\documentclass[12pt]{iopart}

\usepackage[latin1]{inputenc}
\usepackage[T1]{fontenc}

\usepackage{graphics}
\usepackage{graphics}
\usepackage{epsfig}
\usepackage{subfigure}
\usepackage{wrapfig}
\usepackage{iopams, latexsym, amscd, amstext,  setstack, amsthm}
\usepackage{geometry}
\usepackage{pifont}

\usepackage{enumerate}

\newtheorem{thm}{\bf Theorem}[section]
\newtheorem{prop}[thm]{\bf Proposition}
\newtheorem{corol}[thm]{\bf Corollary}
\newtheorem{defin}[thm]{\textsl{\bf Definition}{}}
\newtheorem{lemma}[thm]{\bf Lemma}

\newtheorem{rem}[thm]{\bf Remark}

\newcommand{\Span}{\mathrm{Span}}
\newcommand{\Supp}{\mathrm{Supp}}
\newcommand{\im}{\mathrm{im}}

\begin{document}

\title[Clean POVMs for qubits]{Clean positive operator valued measures for qubits and similar cases}
\author{Jonas Kahn$^1$}
\address{$^1$ Universit\' e Paris-Sud 11, D\' epartement de Math\' ematiques,
B\^{a}t 425, 91405 Orsay Cedex, France}
\ead{jonas.kahn@math.u-psud.fr}


\begin{abstract}
  In a recent paper \cite{BAKPW} was defined a notion of clean positive
operator valued measures (POVMs). We here
characterize which POVMs are clean in some class that we call quasi-qubit
POVMs, namely POVMs whose elements are all rank-one or full-rank. We give an
algorithm to check whether a given quasi-qubit POVM satisfies to this
condition. We describe explicitly all the POVMs that are clean for the qubit. On the
way we give a sufficient condition for a general POVM to be clean.
\end{abstract}
\pacs{03.65.-w}
\ams{47L05, 81T05}
\submitto{\JPA}

\noindent {\em Keywords and phrases\/} Quantum measurements, optimization of
measurements, positive operator valued measures, completely positive
maps, channels.

\maketitle

\section{Introduction}

  The laws of quantum mechanics impose restrictions on what measurements can be
carried out on a quantum system. All the possible measurements can be described mathematically by ``positive operator-valued measures'', POVMs
for short. Apart from measuring a state, we  can also transform it  via a quantum
channel. Now suppose we have at our disposal a POVM {\bf P} and a channel $\mathcal{E}$. We may first send our state through $ \mathcal{E}$ and then
feed the transformed state in our measurement apparatus {\bf P}. This
procedure is a new measurement procedure, and can therefore be encoded by a
POVM {\bf Q}. Now transforming the state with $ \mathcal{E}$ can be seen as a
kind of noise on the POVM {\bf P}. We may then view {\bf Q} as a disturbed
version of {\bf P}, and we  say that {\bf P} is \emph{cleaner} than {\bf
Q}. Now, what are the maximal elements for this order relation?

  The order relation ``cleaner than'' has been introduced in a recent article
of Buscemi \emph{et al.} \cite{BAKPW}. Herein they look at which POVMs can be
obtained from another, either by pre-processing (the situation we just
described, where we first send our state through  a channel) or by
classical post-processing of the data. Especially, they try to find which
POVMs are biggest for these order relations (in the former case, 
the POVM is said to be  \emph{clean}; there is no ``extrinsic'' noise). For pre-processing they get a number
of partial answers. One of those is that a POVM on a $d$-dimensional space with
$n$ outcomes, with $n\leq d$, is clean if and only if it is an observable. They
do not get a complete classification, though. 

  The object of the present article is to characterize which POVMs are clean
in a special class of measurements. Namely, we are interested in
POVMs such that all their elements (see definition below) are either full-rank
or rank-one. We  call these POVMs \emph{quasi-qubit POVMs}. Notice
that all the POVMs for qubits satisfy to this condition. 

  On the way we  prove a sufficient condition for a POVM to be clean, that
is usable also for POVMs that are not quasi-qubit.

  It turns out that cleanness for quasi-qubit POVMs can  be  read on the span of the rank-one elements.  Moreover,if a
(non necessarily quasi-qubit) POVM is cleaner than a clean quasi-qubit POVM, the latter
was in fact obtained by a channel that is a unitary transform. In other words, for quasi-qubit POVMs, cleanness-equivalence is unitary equivalence.
 
  We  give an algorithm to check whether a quasi-qubit POVM is clean
or not. This algorithm may be the main contribution of the article, as almost all the following theorems can be summed up by saying the algorithm is valid. 
 
  In the end we  apply these results to the qubit, for which all
POVMs are quasi-qubit. We are then left with a very explicit characterization
of clean POVMs for qubits.

  Section \ref{notations} gives precise definitions of all the objects we cited
in this introduction. 

We define the algorithm, give heuristically the main ideas and define the important notion  ``totally determined'' (Definition \ref{totdet}) in Section \ref{ialg}.

Section \ref{suff} gives a sufficient condition for a POVM to be clean,
namely that the supports of the elements of the POVM ``totally determine''
the space (see Definition \ref{totdet}). We use this condition to show that
when the algorithm exits with a positive result, the quasi-qubit POVM is really
clean.

  Section \ref{necessary} proves that the above sufficient condition is in fact
necessary for quasi-qubit POVMs. It checks that when the algorithm exits with a
negative result, the POVM is truly not clean.

  Section \ref{summary} gathers the results relative to quasi-qubit POVMs in
Theorem \ref{gather} and deals with the qubit case in Corollary \ref{qubit}.

  Ultimately section \ref{outlook} gives a very rough idea for making explicit  more
explicit the sufficient condition for a POVM
to be clean we have
given in section \ref{suff}.

\smallskip

  If one wishes to look for the results of this paper without bothering with the
technical proofs, the best would be to read the algorithm of section \ref{ialg}
and then
to read Theorem \ref{gather} and Corollary \ref{qubit}. You would also need
Lemma \ref{determination} that you could use as a definition of ``totally
determined'' if you are only interested in quasi-qubit POVMs. 

If you also want the supplementary results that apply to other POVMs, further
read Definitions \ref{corresponding} and \ref{totdet}, and Theorem \ref{suffcond}.

\section{Definitions and notations}
\label{notations}

  We consider POVMs  on a Hilbert space $ \mathcal{H}$ of dimension $d\geq 2$. 
Dimension $2$ is the qubit case. The set $\{|e_i\rangle\}_{1\leq i \leq d}$
will be an orthonormal basis of $ \mathcal{H}$. If $ \mathcal{V}$ is a subspace
of $ \mathcal{H}$ then $ \mathcal{V}^{\bot}$ is the subspace orthogonal to $
\mathcal{V}$ in $ \mathcal{H}$. If we are given vectors $\{v_i\}_{i\in I}$, we
denote by $\Span(v_i, i\in I)$ the space they generate.
The set of operators on $
\mathcal{H}$ is denoted by $ \mathcal{B}( \mathcal{H})$. 

A POVM {\bf P} (with finite
outcomes, case to which we restrict) is a set $\{P_i\}_{i\in I}$ of \emph{non-negative} operators on $
\mathcal{H}$, with $I$ finite, such that $\sum_{i \in I} P_i = \mathbf{1}$. The
$P_i$ are called \emph{POVM elements}. We  write $\Supp(P_i)$ for the
support of this element. This support is defined by its orthogonal. The set of $|\phi\rangle\in \Supp(P_i)^{\bot}$ is exactly the set of $|\phi\rangle$ such that $\langle \phi|P_i|\phi\rangle = 0$. The rank of a POVM element is its rank
as an operator. In particular, rank-one elements are of the form $\lambda_i
|\psi_i\rangle\langle\psi_i|$ and full-rank POVMs are invertible. Special
cases of POVMs are \emph{rank-one POVMs}, that is POVMs whose elements are all
rank-one, and \emph{full-rank POVMs}, that is POVMs whose elements are all
full-rank. We are
especially interested in a class of POVMs that includes both:
\begin{defin}\emph{Quasi-qubits POVMs}

  A POVM {\bf P} is a \emph{quasi-qubit} POVM if all its elements $P_i$ are
either full-rank or rank-one.

Similarly, we shall speak of \emph{strict quasi-qubit} POVMs for quasi-qubit
POVMs which are neither rank-one nor full-rank.
\end{defin}

A channel $ \mathcal{E}$ is a completely positive identity-preserving map on $ \mathcal{B}(
\mathcal{H})$ the set of bounded operators on $ \mathcal{H}$ (in this paper,
channels are always intended as going from  $  \mathcal{B}(
\mathcal{H})$  to the \emph{same}  $  \mathcal{B}(
\mathcal{H})$).  As a remark, this implies that the subspace of self-adjoint operators $
\mathcal{B}_{sa}( \mathcal{H})$ is stable by $ \mathcal{E}$. We know we can
write it using Kraus decomposition \cite{LivreKraus}, that is we can find a finite number of
operators $R_{\alpha}\in \mathcal{B}( \mathcal{H})$ such that
\begin{equation}
\label{Kd}
\mathcal{E}(A) = \sum_{\alpha} R_{\alpha}^* A R_{\alpha}, \quad \mathrm{with\ \ }
\sum_{\alpha} R_{\alpha}^* R_{\alpha} = \mathbf{1}.
\end{equation}
Here the star is the adjoint. 

We shall write $ \mathcal{E}= \{R_{\alpha}\}_{\alpha}$. This decomposition is
not unique. 

Using the channel $ \mathcal{E}$ before the measurement {\bf P}
is the same as using the POVM $\mathbf{Q} = \mathcal{E}(\mathbf{P})$ defined by
its POVM elements $Q_i= \mathcal{E}(P_i)$.

\begin{defin}
A POVM {\bf P} is \emph{cleaner} than a POVM {\bf Q} if and only if there
exists a channel $ \mathcal{E}$ such that $ \mathcal{E}(\mathbf{P})=
\mathbf{Q}$. We shall also write $\mathbf{P}\succ\mathbf{Q}$.
\end{defin}

\begin{defin}\emph{Clean POVM}

  A POVM {\bf P} is clean if and only if, for any {\bf Q} such that
$\mathbf{Q}\succ\mathbf{P}$, then $\mathbf{P}\succ\mathbf{Q}$ also holds.
\end{defin}

We shall further say that two POVMs are cleanness-equivalent if   both
$\mathbf{Q}\succ\mathbf{P}$ and $\mathbf{P}\succ\mathbf{Q}$ hold. A special
case of this (but not the general case, as proved in \cite{BAKPW}) is
\emph{unitary  equivalence}, when there is a unitary operator $U$ such that for
any $i\in I$, we have $U P_i U^* = Q_i$.

\section{ Algorithm and Ideas}
\label{ialg}

\subsection{Algorithm}

We propose the following algorithm to check whether a quasi-qubit POVM {\bf P} is clean
or not.

\begin{enumerate}
{\item \label{isr1}
We check whether {\bf P} is rank-one. If it is, exit with result ``\emph{$\mathbf{P}$ is clean}''. Otherwise:
}
{\item 
\label{basis}
Write the rank-one elements $P_i=\lambda_i
|\psi_i\rangle\langle\psi_i|$ for $1 \leq i \leq n$. Check whether these $|\psi_i\rangle$ generate $ \mathcal{H}$. If not, exit with result ``$\mathbf{P}$ \emph{is not clean}''.
Else:}
{\item 
\label{init}
We can find a basis of $\mathcal{H}$ as a subset of those $|\psi_i\rangle$. We assume that this basis consists of $|\psi_i\rangle$ for
$1\leq i\leq d$.  We define a variable  $ \emph{C}= \{V_j\}_{j\in J}$, consisting in a collection of subspaces whose direct sum is the
Hilbert space $
\mathcal{H}=\bigoplus_j V_j$. We initialize $C$ with $V_i = \Span(|\psi_i\rangle)$ for
$1\leq i\leq d$.
}
{\item For $i$ from $d+1$ to $n$, do:
}
{\item
Write $|\psi_i\rangle = \sum_j v_j$ with $v_j \in V_j$. Call $J(i)=\{j|
v_j\neq 0\}$.
}
{\item
\label{update}
Update $\{V_j\}$: Suppress all $V_j$ for $j \in J(i)$. Add $V_i=\bigoplus_{j\in
J(i)}V_j $.
}
{\item
\label{cleanexit}
Check whether $C = \{\mathcal{H}\}$. If so,  exit with result ``\emph{{\bf P} is clean}''. Otherwise:
}
{\item End of the ``For'' loop.
}
{\item
\label{out}
Exit with result ``\emph{{\bf P} is not clean}''.
}
\end{enumerate}

Notice that the algorithm terminates: every stage is finite and we enter the loop a finite number of times.

\subsection{Heuristics: what the algorithm really tests}
\label{heuristics}

In the Kraus decomposition \eref{Kd}, each of the terms $R_{\alpha}^* A R_{\alpha}$ is non-negative if $A$ is non-negative, so that $\mathcal{E}(A) \geq R_{\alpha}^* A R_{\alpha}$ for any $\alpha$. Hence if $\mathcal{E}(\mathbf{Q}) = \mathbf{P}$, then $R_{\alpha}^* Q_e R_{\alpha}$ must have support included in $\Supp(P_e)$ for all $\alpha$ and $e\in E$.

The central idea of the paper is the following: the condition $\Supp(R_{\alpha}^* Q_e R_{\alpha}) \subset \Supp(P_e)$ yields $d - \dim(\Supp(P_e))$ homogeneous linear equations on the matrix entries of $R_{\alpha}$, where you should remember that $d = \dim(\mathcal{H})$. Now $R_{\alpha}$ is determined up to a constant by $d^2 -1$ homogeneous independent linear equations. In such a case, the additional condition $\sum R_{\alpha}^* R_{\alpha} = \mathbf{1}$ yields all $R_{\alpha}$ are proportional to the same unitary $U$, so that the channel $\mathcal{E}$ is unitary, and $\mathbf{P} \succ \mathbf{Q}$.

There is still one difficulty: the equations mentioned above depend not only on $\mathbf{P}$, but also on $\mathbf{Q}$. We would then like conditions on the supports of $P_e$ such that the system of equations mentioned above is at least of rank $d^2 - 1$ for all $\mathbf{Q}$. We formalize this requirement with the following definitions.

\begin{defin}\emph{Corresponding}
\label{corresponding}

  Let $ \mathcal{V}$ be a Hilbert space and $\{F_i\}_{i\in I}$ a collection of
subspaces of $ \mathcal{V}$. Let $\{v_i\}_{i\in I}$ be a collection of vectors of $\mathcal{V}$. This set of vectors \emph{corresponds to} $\{F_i\}_{i\in I}$ if for any $i \in I $, there is a linear transform  $R_i$ such that $R_i(v_i) \neq 0$ and, for all $j\in
I$, the transform is taking $v_j$ within $F_j$, that is $R_i(v_j)\in F_j$. 

In the text, we usually drop the reference to  $\{F_i\}_{i\in I}$ and write that the  $\{v_i\}_{i\in I}$ are a \emph{corresponding} collection of vectors.
\end{defin}

\begin{defin}\emph{Totally determined}
\label{totdet}

Let $ \mathcal{V}$ be a Hilbert space and $\{F_i\}_{i\in I}$ a collection of
subspaces of $ \mathcal{V}$.

  If for all corresponding collections of vectors $\{v_i\}_{i\in I}$ there is
only one (up to a complex multiplicative constant) linear transform $R$ such
that $R(v_i)\in F_i$ for all $i \in I$, we say that $ \mathcal{V}$ is
\emph{totally determined} by $\{F_i\}_{i\in I}$, or alternatively that
$\{F_i\}_{i\in I}$ \emph{totally determines} $ \mathcal{V}$.

If $F_i$ is one-dimensional with support vector $w_i$, this means there is only
one $R$ such that $R(v_i)$ is colinear to $w_i$ for all $i\in I$.
\end{defin}

What the algorithm does is checking that a quasi-qubit POVM {\bf P}  is rank-one (stage (\ref{isr1})), or that {\bf P} totally determines $\mathcal{H}$. 

More precisely, Proposition \ref{stage6} states that each of the $V_j$ belonging to  $C$ (appearing at stage (\ref{init}) and updated at stage (\ref{update})) is totally determined by the $|\psi_i\rangle$ such that $|\psi_i\rangle \in V_j$. When the algorithm exits at stage (\ref{cleanexit}), then $ C = \{\mathcal{H}\} $, so $\mathcal{H}$ is totally determined. If the algorithm does not exit at stage (\ref{cleanexit}), on the other hand, then $C$ has at least two elements at the last stage, and each $|\psi_i\rangle$ is included in one of those two elements, which entails, from Lemma \ref{determination}, that $\{\Supp(P_e)\}$ does not totally determine $\mathcal{H}$.

The equivalence with cleanness for quasi-qubit POVMs is still needed to get validity of the algorithm. This equivalence stems from Theorem \ref{suffcond} and Theorem \ref{reciproque}. The former is the sufficient condition, for any POVM, not necessarily quasi-qubit. We have given the intuition for this theorem at the beginning of the section. Complementarily, Theorem \ref{reciproque} states that a strict quasi-qubit POVM is not clean if its supports do not totally determine $\mathcal{H}$.

The proof of Theorem \ref{reciproque} features the last important idea of the paper. A channel $\mathcal{E}$ which is near enough the identity may be inverted as a positive map on $\mathcal{B}(\mathcal{H})$, even though $\mathcal{E}^{-1}$ is not a channel. Now if we denote $\mathbf{Q} = \mathcal{E}^{-1} (\mathbf{P})$, we have          $\mathcal{E}(\mathbf{Q}) = \mathbf{P}$. We are then left with two questions: is {\bf Q} a POVM, and can we find a channel $\mathcal{F}$ such that $\mathcal{F}(\mathbf{P}) = \mathbf{Q}$?

The main possible obstacle to {\bf Q} being a POVM is the need for each of the $Q_i$ to be non-negative. Now, if $\mathcal{E}$ is near enough the identity, if $P_i$ was full-rank, then $Q_i$ is still full-rank non-negative. The remaining case is $Q_i = \mathcal{E}^{-1}(P_i) = \lambda_i \mathcal{E}^{-1}\left( |\psi_i\rangle \langle \psi_i| \right) $. Now, we shall see that we may use the set of subspaces $C = \{V_j\}$ given by the algorithm to build channels ensuring that these $Q_i$ are still rank-one non-negative matrices. Furthermore, these $Q_i$ will have a bigger first eigenvalue than $P_i$, so that we are sure {\bf Q} is strictly cleaner than {\bf P}, as channels are spectrum-width decreasing (see Lemma \ref{spectrum}).

\medskip

We now turn to the fully rigorous treatment.

\section{Sufficient condition}
\label{suff}

We start by proving the following theorem, announced in the previous section.

\begin{thm}
\label{suffcond}
  If the supports $\{\Supp(P_i)\}_{i\in I}$ of the elements $P_i$ of a POVM {\bf
P} totally determine $ \mathcal{H}$, then {\bf P} is clean and any
cleanness-equivalent POVM {\bf Q} is in fact unitarily equivalent to {\bf P}.
\end{thm}

\begin{proof} It is enough to prove that if $\mathbf{ Q} \succ \mathbf{P}$, then {\bf Q} is unitarily equivalent to {\bf P}.

 Let {\bf Q} be a POVM and $
\mathcal{E}=\{R_{\alpha}\}_{\alpha}$ a channel such
that $ \mathcal{E}(\mathbf{Q})= \mathbf{P}$.

For all $i \in I$, we may write $Q_i = \sum_{k} \mu_{i,k} |\phi_i^k\rangle\langle\phi_i^k|$. Then we have $P_i = \sum_{\alpha}  \sum_{k}  \mu_{i,k} R_{\alpha}^* |\phi_i^k\rangle\langle\phi_i^k| R_{\alpha}$. Now $ \mu_{i,k} R_{\alpha}^* |\phi_i^k\rangle\langle\phi_i^k| R_{\alpha} \geq \mathbf{0}$ for all $k$ and $\alpha$, and consequently  $ \mu_{i,k} R_{\alpha}^* |\phi_i^k\rangle\langle\phi_i^k| R_{\alpha} \leq P_i$. Hence $R_{\alpha}^* |\phi_i^k\rangle \in \Supp(P_i)$.

Moreover $P_i$ is nonzero. So that there is at least one $k(i)$ and one $ \alpha(i)$ for each $i$ such that $R_{\alpha}^* |\phi_i^{k(i)}\rangle $  is nonzero. Thus 
$\{\phi_i^{k(i)}\}_{i\in I}$ corresponds to  $\{\Supp(P_i)\}_{i\in I}$. As  $\{\Supp(P_i)\}_{i\in I}$ totally determines $ \mathcal{H}$, there is only one $R$, up to a constant, such that $ R |\phi_i^{k(i)}\rangle \in \Supp(P_i)$ for all $i$. So that $R_{\alpha} = c(\alpha) R$ for all $\alpha$. Since $\sum_{\alpha} R_{\alpha}^* R_{\alpha} = \mathbf{1}$, there is a constant such that $\lambda R_1 $ is unitary, and $\mathcal{E} = \{\lambda R_1\}$. So that {\bf P} and {\bf Q} are unitarily equivalent.

\end{proof}

Before proving in Theorem \ref{stage6} that ``when the algorithm exits at stage (\ref{cleanexit}), then the supports of the POVM {\bf P} totally determine $\mathcal{H}$'', we need a few more tools.

We first need the notion of \emph{projective frame}. Indeed, in the algorithm, we are dealing with supports of rank-one POVMs, that is essentially projective lines. And we want them to totally determine the space, that is essentially fix it. Projective frames are the most basic mathematical object meeting these requirements. We redefine them here, and reprove what basic properties we need; further information on projective frames may be found in most geometry or algebra textbooks, e.g. \cite{projfram}.

\begin{defin}
  A \emph{projective frame} $\{v_i\}_{1\leq i \leq d+1}$ of a vector space $ \mathcal{V}$ is a set of $(\dim(
\mathcal{V}) +1) $ vectors in general position, that is, such that any subset of $ \dim( \mathcal{V})$
vectors is a basis of $ \mathcal{V}$.
\end{defin}

\begin{rem}
\label{projframe}
  Equivalently we may say that $\{v_i\}_{1\leq i\leq n}$ is a basis of $
\mathcal{V}$ and $ v_{d+1} = \sum_{i=1}^n c_i v_i$ with all $c_i\neq 0$. 
\end{rem}

\begin{prop}
\label{projtot}
 A projective frame $\Psi=\{e_i\}_{1\leq i\leq (n+1)}$of $ \mathcal{V}$ totally determines $ \mathcal{V}$.
\end{prop}
\begin{proof}

First we prove that if  $\Phi=\{v_i\}_{1\leq i\leq (n+1)}$ is not a projective frame, the set of vectors  $\{v_i\}_{1\leq i\leq (n+1)}$ does not correspond to $\Psi$. Indeed, as $\Phi$ is not a projective frame,  we may find $n$ vectors, say the $n$ first, such that $\sum_{i=1}^n a_i v_i = 0$ with
at least one $a_i$ non-zero, say $a_1$. Then for any $R$ such that $R(v_i)$ is colinear to
$e_i$ for all $i$, we still have $\sum_{i=1}^n a_i R(v_i) = 0$. As
$\{e_i\}_{1\leq i\leq n}$ is a basis, $a_i R(v_i)= 0$ for all $i$, so that $R(v_1)=0$. Hence  $\{v_i\}_{1\leq i \leq n+1}$ does not correspond to $\{e_i\}_{1\leq i \leq n+1}$. 

Let now $\Phi=\{v_i\}_{1\leq i\leq (n+1)}$ be corresponding to $\Psi$. Notably, this implies that  $\Phi$ is a projective frame. Furthermore, there
is a nonzero linear transform $R$ such that $R(v_i)$ is colinear to $e_i$ for all $i$.  We must show that $R$ is unique up to a constant.

We know that $\{e_i\}_{1\leq i \leq n}$ and $\{v_i\}_{1\leq i \leq n}$ are both
bases of $ \mathcal{V} $. Hence there is a unique transfer matrix  $X$ from the
latter basis to the former. Since $R(v_i) = D_i e_i$ for some $D_i$, we know that $R$ is of the form $D X$ where $D$ is a
diagonal matrix with diagonal values $D_i$. 

We still have not used our $(n+1)$th condition. We are  dealing with projective
frames, so that  $ e_{n+1} = \sum_{i=1}^n b_i e_i$ and $ v_{n+1} = \sum_{i=1}^n c_i
v_i$ with all $b_i$ and $c_i$ non-zero. Now $R(v_{n+1})= \sum_{i=1}^n c_i
R(v_i) = \sum_{i=1}^n c_i D_i e_i$, so that $c_i D_i / b_i$ must be independent
on $i$ and $D$ and hence $R$ is fixed up to a complex multiplicative constant.

\end{proof}

We now turn to a few observations about totally determined spaces.

\begin{rem}
\label{invertible}
If $\{F_i\}_{i \in I}$ totally determines $\mathcal{H}$, and if $\{v_i\}_{i\in I}$ corresponds to $\{F_i\}$, then the up to a constant unique nonzero $R$ such that $R v_i \in F_i$ for all $i\in I$ is invertible.
\end{rem}
\begin{proof}
Let us define $\Pi_{(\ker R)^\perp}$ the projector on the orthogonal of the kernel  of $R$ along its kernel, and $\Pi_{\ker R}$ the projector on the kernel of $R$ along $(\ker R)^{\perp}$. We have $R = R \Pi_{(\ker R)^{\perp}} $, so that $R \Pi_{(\ker R)^\perp} v_i = R v_i$. Thus $\{\Pi_{(\ker R)^\perp} v_i\}_{i\in I}$ is corresponding to $\{F_i\}_{i\in I}$. On the other hand, $\Pi_{\ker R} \Pi_{(\ker R)^\perp} = 0$, so that $(R + \Pi_{\ker R}) (\Pi_{(\ker R)^\perp} v_i) = R (\Pi_{(\ker R)^\perp} v_i) \in F_i$. As $\{\Pi_{(\ker R)^\perp}\}$ is corresponding to $\{F_i\}$, the latter equality implies that $R$ is proportional to $(R + \Pi_{\ker R})$. This is only possible if $\Pi_{\ker R} = 0$. Hence $R$ is invertible. 
\end{proof}

\begin{rem}
\label{subset}
If $\{v_l\}_{l\in I\cup J}$ is corresponding to $\{F_l\}_{l\in I \cup J}$, then $\{v_i\}_{i\in I}$ (resp. $\{v_j\}_{j\in J}$) is corresponding to $\{F_i\}_{i\in I}$ (resp. $\{F_j\}_{j\in J}$.
\end{rem}
\begin{proof}
The set $I$ is a subset of $I \cup J$, thus, for all $i\in I$, there is an $R_i$ such that $R_i v_i \neq 0$ and $R_i v_l \in F_l$ for all $l\in I\cup J$. \emph{A fortiori} $R_i v_k \in F_k$ for all $k\in I$. Hence $\{v_i\}_{i\in I}$ is corresponding to $\{F_i\}_{i\in I}$. The same proof yields the result for $J$.
\end{proof}

\begin{rem}
\label{all}
If $\{v_i\}_{i\in I}$ is corresponding to $\{F_i\}_{i\in I}$, then there exists $R$ such that $R v_i \in F_i$ and $R v_i \neq 0$ for all $i$ simultaneously.
\end{rem}
\begin{proof}
By the definition of ``corresponding to'', we have a set $\{R_i\}_{i\in I}$ of transforms such that $R_i v_i  \neq 0$ and $R_i v_j \in F_j$ for all $j\in I$. Now, for any set of coefficients $\{a_i\}_{i\in I}$ the matrix $R = \sum_i a_i R_i$ fulfils $R v_i \in F_i$ for all $i$. If we choose appropriately $\{a_i\}$ we also have $R v_i \neq 0$. For example, we may write all the $R_i v_i$ in the same basis, take note of all coordinates, and choose the $a_i$ as any real numbers algebraically independent of those coordinates.
\end{proof}

\begin{lemma}
\label{addition}
  If $ \mathcal{V}$ and $ \mathcal{W}$ are both totally determined by sets of
subspaces $\{F_i\}_{i\in I} $ and $ \{ F_j\}_{j\in J}$ and if $ \mathcal{V}$
and $ \mathcal{W}$ intersect (apart from the null vector), then their sum $
\mathcal{U} = \mathcal{V} + \mathcal{W}$ is totally determined by $\{F_l\}_{l\in I\cup J}$.
\end{lemma}

\begin{proof}

Let $\{u_l\}_{l\in I\cup J}$ vectors of $ \mathcal{U}$
correspond to $\{F_l\}_{l\in I\cup J}$.
In other words, there is an $R^*$ such that $R^* u_l \in F_l$ for all $l \in I\cup J$. By Remark \ref{all}, we may assume that $R^*u_l \neq 0$ for all $l$. We must show that $R^*$ is unique up  to a constant. Notice that the restriction $R^* u_l \neq 0$ does not play a role: if we find another $R$ non proportional to $R^*$, such that $R u_l \in F_l$ for all $l$, then $R^* + a R$ for appropriate $a$ also fulfils $0 \neq (R^* + a R) u_l \in F_l$ for all $l$, and is not proportional to $R^*$.

\smallskip

We need a few notations. 
First, we consider the space  $ \mathcal{X}= \mathcal{V}\cap \mathcal{W} $. We also define $
\mathcal{Y}$ by $ \mathcal{V}= \mathcal{Y} \oplus
\mathcal{X}$ and $\mathcal{Z}$ by $ \mathcal{W} = \mathcal{Z} \oplus \mathcal{X}$. We  write $I_{\mathcal V}$ and $I_{\mathcal W}$ for the natural inclusions of $\mathcal V$ and $\mathcal W$ in $\mathcal U$. 
We also denote  by  $\Pi_{\mathcal{V}}$ for the projector on $\mathcal{V}$ along $\mathcal{Z}$, by $\Pi_{\mathcal{W}}$  the projector on $\mathcal{W}$ along $\mathcal{Y}$, and by $\Pi_{\mathcal X}$  the projector on ${\mathcal X}$ along $\mathcal Y + \mathcal Z$. 

Please be aware that we do not define $\Pi_{\mathcal V}$ and $\Pi_{\mathcal W}$ as endomorphisms of $\mathcal U$, but as applications from $\mathcal U$ to $\mathcal V$ and $\mathcal W$, respectively. The corresponding endomorphisms are $I_{\mathcal V} \Pi_{\mathcal  V}$ and $I_{\mathcal W} \Pi_{\mathcal W}$.

\smallskip

As a first step, we show that $I_{\mathcal V} \Pi_{\mathcal V} R^*$ is unique up to a constant.

 The rank of $I_{\mathcal V} \Pi_{\mathcal V} R^*$ is at most $\dim(\mathcal V)$, so we can factorize it by $\mathcal V$: there exists two linear applications $L_{\mathcal V}^{\mathcal U}$ from $\mathcal U$ to $\mathcal V$ and 
$L_{\mathcal U}^{\mathcal V}$ from $\mathcal V$ to $\mathcal U$, such that 
$I_{\mathcal V} \Pi_{\mathcal V} R^* L^{\mathcal V}_{\mathcal U} L^{\mathcal U}_{\mathcal V} = I_{\mathcal V} \Pi_{\mathcal V} R^*$.

Now for all $i\in I$, we have $R^* u_i \in F_i \subset \mathcal V$, so that $R^* u_i = I_{\mathcal V} \Pi_{\mathcal V} R^* u_i = I_{\mathcal V} \Pi_{\mathcal V} R^* L^{\mathcal V}_{\mathcal U} L^{\mathcal U}_{\mathcal V} u_i$, so that for all $i\in I$ we have the inclusion $0 \neq  (\Pi_{\mathcal V} R^* L^{\mathcal V}_{\mathcal U}) (L^{\mathcal U}_{\mathcal V} u_i) \in F_i$, where we have used $R^* u_l \neq 0$.. Thus $\{L^{\mathcal U}_{\mathcal V} u_i\}_{i\in I}$ is corresponding to $\{F_i\}_{i\in I}$. On the other hand, we know that $\{F_i\}_{i\in I}$ totally determine $\mathcal V$. Hence there is a nonzero constant $\lambda_{\mathcal V}$, and a  $R_{\mathcal V}$ depending only on $\{F_i\}_{i\in I}$, such that $\Pi_{\mathcal V} R^* L^{\mathcal V}_{\mathcal U} = \lambda_{\mathcal V} R_{\mathcal V}$. Moreover, by Remark \ref{invertible}, $R_{\mathcal V}$ is invertible. So that finally $ I_{\mathcal V} \Pi_{\mathcal V} R^* = \lambda_{\mathcal V} I_{\mathcal V} R_{\mathcal V} L^{\mathcal U}_{\mathcal V}$, with image $\im( \lambda_{\mathcal V} I_{\mathcal V} R_{\mathcal V} L^{\mathcal U}_{\mathcal V}) = \mathcal V$. 
Replacing $\mathcal V$ with $\mathcal W$, we get similarly  $ I_{\mathcal W} \Pi_{\mathcal W} R^* = \lambda_{\mathcal W} I_{\mathcal W} R_{\mathcal W} L^{\mathcal U}_{\mathcal W}$.

\smallskip

The last step consists in proving that the two constants $\lambda_{\mathcal V}$ and $\lambda_{\mathcal W}$ are proportional, independently of $R^*$. 
 
We notice that $\Pi_{\mathcal X} I_{\mathcal V} \Pi_{\mathcal V} = \Pi_{\mathcal X}  = \Pi_{\mathcal X} I_{\mathcal W} \Pi_{\mathcal W}$. Hence $\lambda_{\mathcal V} \Pi_{\mathcal X} I_{\mathcal V} R_{\mathcal V} L^{\mathcal U}_{\mathcal V} = \lambda_{\mathcal W} \Pi_{\mathcal X} I_{\mathcal W} R_{\mathcal W} L^{\mathcal U}_{\mathcal W}$. As $\mathcal X \subset \mathcal V$ and $\im(\lambda_{\mathcal V} I_{\mathcal V} R_{\mathcal V} L^{\mathcal U}_{\mathcal V}) = \mathcal V$, we know that $  \lambda_{\mathcal V} \Pi_{\mathcal X} I_{\mathcal V} R_{\mathcal V} L^{\mathcal U}_{\mathcal V} \neq 0$. The equality $\lambda_{\mathcal V} \Pi_{\mathcal X} I_{\mathcal V} R_{\mathcal V} L^{\mathcal U}_{\mathcal V} = \lambda_{\mathcal W} \Pi_{\mathcal X} I_{\mathcal W} R_{\mathcal W} L^{\mathcal U}_{\mathcal W}$ then yields the proportionality of $\lambda_{\mathcal W}$ and $\lambda_{\mathcal V}$. 

We conclude by recalling that $\mathcal V + \mathcal W = \mathcal U$, so that knowing both $I_{\mathcal V} \Pi_{\mathcal V} R^*$ and $I_{\mathcal W} \Pi_{\mathcal W} R^*$ is equivalent to knowing $R^*$. As our only free parameter is the multiplicative constant $\lambda_{\mathcal V}$, we have proved uniqueness of $R^*$, up to a constant.

\end{proof}

Lemma \ref{addition} and Proposition \ref{projtot} are the two ingredients for proving the following proposition, central for the validity of the algorithm. 

\begin{prop}
\label{stage6}
In the algorithm, the spaces in the set $C = \{V_j\}_{j\in J}$ are always totally determined by the supports $K(j) =  \{\Span(|\psi_i\rangle) : |\psi_i\rangle \in V_j\}$ of the one-dimensional POVM elements they contain.
\end{prop}

\begin{proof}
We prove the proposition by induction on the stronger property $Prop =$ `` all $V_j$ are totally determined by  $K(j)$, and they are spanned by vectors of the initial basis, that is, they are of the form $\Span(|\psi_i\rangle : i \in I(j))$, where $I(j)$ is a subset of $\{1,\dots, d\}$''.

\smallskip

\emph{Initialization:} 
We initialize $C$ at step (\ref{init}). At this stage $V_j$ is defined for $j\in \{1,\dots, d\}$ by $V_j = \Span(|\psi_j\rangle) $. So that on the one hand $V_j$ is of the form $ \Span(|\psi_i\rangle : i \in I(j))$, where $I(j)$ is a subset of $\{1,\dots, d\}$, and on the other hand $V_j$ is totally determined by $K(j)$, as it is one-dimensional and $|\psi_j\rangle$ is nonzero.

\smallskip

\emph{Update:}
We update $C$ at stage (\ref{update}). We must prove that $V_i = \bigoplus_{j\in J(i)} V_j$ still fulfils $Prop$. 

For one thing, the space $V_i$ is a sum of spaces of the form  $ \Span(|\psi_i\rangle : i \in I(j))$, where $I(j)$ is a subset of $\{1,\dots, d\}$, hence $V_i$ is also of this form with $I(i) = \bigcup_{j\in J(i)} I(j)$.

Now let us consider the set $I_{int} = \left\{j: j \in \{1\dots d\}, \langle \psi_i | \psi_j \rangle \neq 0  \right\}$, and the space $V_{int} = \Span(|\psi_j\rangle: j\in I_{int})$. Since the $|\psi_j\rangle$ for $j\in I_{int}$ are part of the initial basis $\{|\psi_j\rangle\}_{1\leq j \leq d}\}$, they are independent. The definition of $I_{int}$ also ensures $|\psi_i\rangle = \sum_{j\in I_{int}} c_j |\psi_j\rangle$ with $j$ nonzero, hence, by Remark (\ref{projframe}), the set $\{|\psi_k\rangle: k = k\in I_{int}\cup \{i\}\}$ is a projective frame of $V_{int}$. So that, by Proposition \ref{projtot}, the space $V_{int}$ is totally determined by $\{|\psi_j\rangle\}_{j\in I_{int}\cup \{i\}}$. We initialize $K_{int} = I_{int} \cup \{i\}$.

Finally, by definition of $J(i)$, we know that $V_{int} \cap V_j  \neq 0$ for all $j \in J(i)$. Both are totally determined, by $K(j)$ and $K_{int}$. Hence by Lemma \ref{addition}, $V_{int} \cup V_j$ is totally determined by $K(j) \cup K_{int}$. We update $V_{int} = V_{int} \cup V_j$ and $K_{int} = K_{int} \cup K(j)$. We iterate the latter step for all $j \in J(i)$ and we end up with $V_{int} = V_i$ totally determined by $\bigcup_{j\in j(i)} K(j) \cup I_{int} \cup \{i\} \subset I(i)$.

\end{proof}

\begin{corol}
\label{clean}
When the algorithm ends at stage (\ref{cleanexit}), the POVM {\bf P} is clean.
\end{corol}
\begin{proof}
The algorithm ends at stage (\ref{cleanexit}) only if $C = \{\mathcal H\}$. By the above proposition, this condition implies that $\mathcal H $ is totally determined by $\{\Span(|\psi_j\rangle): |\psi_j\rangle \in \mathcal H\}$. This amounts at saying that $\mathcal H$ is totally determined by the supports of the POVM elements $P_i$, and we conclude by Theorem \ref{suffcond}.
\end{proof}

 This section aims at giving sufficient conditions for a POVM to be clean, and at proving that one of these conditions is fulfilled if the algorithm exits with result ``{\bf P} is clean''. We thus conclude the section with the case when the algorithm exits at stage (\ref{isr1}). In other words, we must show that a rank-one POVM is clean. Now, this has already been proved as 
 Theorem 11.2 of \cite{BAKPW}:
\begin{thm}
\label{rankone}
\cite{BAKPW}
  If {\bf P} is rank-one, then {\bf Q}$\succ${\bf P} if and only if {\bf P} and {\bf Q}
are unitarily equivalent. Thus, rank-one POVMs are clean.
\end{thm}

For a quasi-qubit POVM {\bf P}, we prove in the following section that {\bf P} is clean only if it  fulfils the  conditions either of Theorem \ref{rankone} or of Theorem \ref{suffcond}.

\section{Necessary condition for quasi-qubit POVMs}
\label{necessary}

This section proves that a clean quasi-qubit POVM either is  rank-one, or the supports of its elements totally determine the space:

\begin{thm}
\label{reciproque}
  A non-rank-one quasi-qubit POVM where  $\{\Supp(P_i)_{i\in I}\}$ does not determine $ \mathcal{H}$
is not clean.
\end{thm}   

We need a few more tools to prove the theorem.

To begin with, we need a way to prove in specific situations that a POVM is not cleaner than another. 
 Using the fact that channels are \emph{spectrum-width
decreasing} is the easiest method. This is Lemma 3.1 of \cite{BAKPW}:

\begin{lemma}
\label{spectrum}
If the minimal (resp.
maximal) eigenvalue of $X$ is denoted $\lambda_m(X)$ (resp.
$\lambda_M(X)$), then $\lambda_m(X)\leq  \lambda_m( \mathcal{E}(X)) \leq
\lambda_M( \mathcal{E}(X))\leq \lambda_M(X)$ for any channel $ \mathcal{E}$.
\end{lemma}

This lemma implies that existence of $\mathbf{Q}\succ\mathbf{P}$ such that for some
$i\in I$, either $\lambda_m(Q_i)<\lambda_m(P_i)$ or
$\lambda_M(Q_i)>\lambda_M(P_i)$ entails that {\bf Q} is strictly
cleaner than {\bf P},  so that {\bf P} is not clean.

We now give a characterization of the fact that $ \mathcal{H}$ is totally
determined by $\{F_j\}_{j\in J}$ when all the $F_j$ are one-dimensional, that
is of when the $F_j$ can be seen as vectors. This characterization applies to
$\{\Supp(P_i)\}_{i\in I}$ for quasi-qubit POVMs, and may be more intuitive than
 Definition \ref{totdet}. Moreover it is more adapted to our strategy of
proof.

\begin{lemma}
\label{determination}
A set of vectors $\{ |\psi_j\rangle\}_{j\in J}$ totally determine the space $
\mathcal{H}$, if and only if, for any two supplementary proper subspaces $ \mathcal{V}$ and $
\mathcal{W}$, there is a $j\in J$ such that $ |\psi_j\rangle\not\in
\mathcal{V}$ and $ |\psi_j\rangle
\not\in \mathcal{W}$.

Moreover, when the algorithm exits with result ``{\bf P} is not clean'', the supports of {\bf P} do not totally determine $\mathcal H$.
\end{lemma}

\begin{proof}

The proof is made of four steps: 
\begin{enumerate}[(a)]
\item{
\label{s0}
For any finite  set of vectors $\{|\psi_j\rangle\}_{j\in J}$, there is a POVM whose supports of the rank-one elements are these vectors.
}
\item {
\label{sa}
if we feed into the algorithm a non-rank-one quasi-qubit POVM whose supports of rank-one elements are the $|\psi_j\rangle$ and if $\{|\psi_j\rangle\}$ does not totally determine $\mathcal H$, then the algorithm exits with result ``{\bf P} is not clean''.}
\item{ 
\label{sb}
 if the algorithm exits with result   ``{\bf P} is not clean'', then we can find two supplementary proper subspaces such that $|\psi_j\rangle \in \mathcal V$ or $|\psi_j\rangle \in \mathcal W$ for all supports of rank-one elements.}
\item{ 
\label{sc}
finding  two supplementary proper subspaces such that $|\psi_j\rangle \in \mathcal V$ or $|\psi_j\rangle \in \mathcal W$ for all $j\in J$ implies that $\{|\psi_j\rangle\}_{j\in J}$ does not totally determine $\mathcal H$.}
\end{enumerate}

The equivalence in the lemma is then proved by contraposition, and the last statement by combining (\ref{sb}) and (\ref{sc}).

\emph{Step (\ref{s0}):} 
A valid example is given by $P_j = \frac1{2\#J} |\psi_j\rangle\langle \psi_j|$ for $j\in J$ and $P_{\#J + 1}= \mathbf{1} - \sum_j P_j$. Indeed the latter element is positive since $\sum_j P_j \leq \frac1{2\#J}  \#J \mathbf{1} = \frac12 \mathbf{1}$.

\emph{Step (\ref{sa}):}  Since the quasi-qubit POVM is assumed not to be rank-one, we do not exit at stage (\ref{isr1}). The only other possible exit with result ``{\bf P} is clean'' is at stage (\ref{cleanexit}). Now the proof of Corollary \ref{clean}  states that the algorithm exits at stage (\ref{cleanexit}) only if the supports of the rank-one elements totally  determine $\mathcal H$. Hence, the algorithm exits with result ``{\bf P} is not clean''.

\emph{Step (\ref{sb}):} 
 Exiting at stage (\ref{basis}) means that the
$|\psi_j\rangle$ do not generate $\mathcal H$.
Then, if  $J = \varnothing$, we may choose any two supplementary proper subspaces $\mathcal V$ and $\mathcal W$. Anyhow $|\psi_j\rangle\in \mathcal V$ for all $j\in J$.
If $J\neq\varnothing$,
then  $ \mathcal{V} = \Span(|\psi_i\rangle, i\in I)$ is a proper subspace of $\mathcal H$.  Since $|\psi_j\rangle \in \mathcal V$ for all $j\in J$, any supplementary subspace $
\mathcal{W}$ of $ \mathcal{V}$ will turn the trick.

If the algorithm does not exit at stage (\ref{basis}), then 
there is a basis included in $\{|\psi_j\rangle\}_{j\in J}$. We assume that it corresponds to $1\leq j\leq d$.

Since the algorithm exits with result, ``{\bf P} is not clean'', it exits at stage (\ref{out}). We 
end the algorithm with a collection $C=\{V_k\}$ of subspaces such that
$\bigoplus_k V_k = \mathcal{H}$. 
 Since we have not exited at stage (\ref{cleanexit}), we know that $C\neq \{\mathcal H\}$. Hence $C$ counts at least two
non-trivial elements. We take  $ \mathcal{V}= V_1$ and $ \mathcal{W} =
\bigoplus_{k\neq1} V_k$.  

 The $V_k$ are direct sums of the original $V_j = \Span(|\psi_j\rangle)$ for $1\leq j\leq d$. Hence, for $1 \leq j\leq d$, either $|\psi_j\rangle \in \mathcal V$ or $|\psi_j\rangle \in \mathcal W$.
 On the other hand if
$|\psi_j\rangle$ is not one
of the original basis vectors, it was used in the ``For'' loop. At the end of
this loop, $C$ was then containing a space $V= \bigoplus_{k\in J(j)}V_k$. And
$|\psi_j\rangle$ was included in this space. This $V$ is then included in one
of the final $V_j$ and a fortiori either in $ \mathcal{V}$ or in $
\mathcal{W}$. We have thus proved that when the algorithm
exits with a negative value we may find two supplementary proper
subspaces $ \mathcal{V}$ and $
\mathcal{W}$ such that for all $i\in I$, either $ |\psi_i\rangle\in
\mathcal{V}$ or $ |\psi_i\rangle
\in \mathcal{W}$. 

\emph{Step (\ref{sc}):}
Since $\mathbf{1} |\psi_j\rangle =  |\psi_j\rangle$ for all $j$, by Definition \ref{corresponding} the set of vectors $\{|\psi_j\rangle\}_{j\in J}$ is corresponding to   the subspaces $\{|\psi_j\rangle\}_{j\in J}$. 
On the other hand,  denoting by $\Pi_{\mathcal{V}}$ the projection on $ \mathcal{V}$ parallel to $ \mathcal{W}$, we get that   $\Pi_{\mathcal{V}}|\psi_j\rangle $ is colinear to
$|\psi_j\rangle$ for all $j\in J$. Moreover $\Pi_{\mathcal V}$ is not proportional to $\mathbf{1}$, so that, by  definition \ref{totdet}, the set of
vectors $\{|\psi_j\rangle\}$ does not totally determine $ \mathcal{H}$.

\end{proof}

Finally, as explained in Section \ref{ialg}, we want to build our cleaner POVMs as $\mathcal E^{-1}(\mathbf P)$ where the channel is inverted as a positive map. We need to know some conditions under which a channel can be inverted. This is the purpose of Lemma \ref{channelsidentity}, for which we need the following norms.

The Hilbert-Schmidt norm on $ \mathcal{B}( \mathcal{H})$ is defined  as  $ \| M \|^2_{HS} =
\Tr(MM^*)$.
Notably, in any
orthogonal basis,
\[
\| M \|^2_{HS} = \sum_{1\leq i,j\leq d} |M_{i,j}|^2.
\]
Moreover $\|M\|_{HS} = \| M^* \|_{HS}$. 

We also define a norm on $ \mathcal{B}( \mathcal{B} ( \mathcal{H} ))$, space to
which the channels belong:
\[
\| \mathcal{O} \|_{1} =  \sup_{\{M| \|M\|_{HS} = 1 \}} \| \mathcal{O}(M) \|_{HS}.
\]

\begin{lemma}
\label{channelsidentity}
If in the Kraus representation of a channel $ \mathcal{E}=\{R_{\alpha}\}$ one
of the $R_{\alpha}$ fulfils
\[ \| \mathbf{1} - R_{\alpha} \|_{HS} \leq \epsilon, \] 
then 
\begin{equation}
\label{bound}
\| \mathbf{1} - \mathcal{E}\|_1 \leq 2(1+\sqrt{d})\epsilon +  2\epsilon^2
= f(\epsilon) \underset{\epsilon\to 0}{\longrightarrow} 0. 
\end{equation}

  As a consequence, if $ f(\epsilon) < 1$, then  $ \mathcal{E}$ is invertible
(as a map on $
\mathcal{B}( \mathcal{H})$) and $ \|\mathcal{E}^{-1} - \mathbf{1}\|_1 \leq
f(\epsilon)/(1-f(\epsilon))$. This inverse lets $ \mathcal{B}_{sa}( \mathcal{H})$
stable.

 This in
turn shows that for any $X\in \mathcal{B}_{sa}( \mathcal{H})$
such that $\lambda_m(X)\geq 0$, the spectrum of the image by the inverse is
bounded through 
\begin{equation}
\label{basspectre}
 \lambda_m(X) - \lambda_M(X)f(\epsilon)\sqrt{d}/(1-f(\epsilon))\leq \lambda_m(
\mathcal{E}^{-1}(X)).
\end{equation} 

So that for all $ X > 0$, when
$\epsilon$ small enough, $ \mathcal{E}^{-1}(X) \geq 0$. 
\end{lemma}
{\bf Remark}: The bound (\ref{bound}) is probably far from sharp, but
sufficient for our needs.
 
\begin{proof} 
Without loss of generality, we assume that 
\[ \| \mathbf{1} - R_{1} \|_{HS} \leq \epsilon. \]

We write $S = R_{1}-\mathbf{1}_{
\mathcal{H}}$ and $ \mathcal{O}= \mathcal{E} -\mathbf{1}_{\mathcal{B}(
\mathcal{H})}$.

Then 
\[
 \mathcal{O}: M \mapsto S^*MS + S^*M + MS + \sum_{\alpha\neq 1}
R_\alpha^*MR_{\alpha}.
\]

And 
\begin{eqnarray*}
\| \mathcal{O}\|_1&  = \sup_{\{M| \|M\|_{HS} = 1 \}} \left\| S^*MS + S^*M + MS +
\sum_{\alpha\neq 1}
R_\alpha^*MR_{\alpha} \right\|_{HS} \\
    & \leq \sup_{\{M| \|M\|_{HS} = 1 \}} \|S^*\| \|M\| \|S\| + \|S^*\| \|M\|
+ \|M\| \|S\| + \sum_{\alpha\neq 1} \|R_\alpha^*\| \|M\| \|R_{\alpha}\| \\
   & = \|S\|_{HS}^2 + 2  \|S\|_{HS} + \sum_{\alpha\neq 1} \|R_{\alpha}\|_{HS}^2.
\end{eqnarray*} 

Now, for one thing, by hypothesis, $\|S\|_{HS}\leq \epsilon$. Furthermore 
\[
\sum_{\alpha\neq 1} \|R_{\alpha}\|_{HS}^2 = \sum_{\alpha\neq 1} \Tr(R_\alpha^*
R_\alpha)  = \Tr(\mathbf{1} - R_1^*R_1) = - \Tr(S^*S + S + S^*).
\]
We finish our proof of (\ref{bound}) with the observation that $-\Tr(S
+ S^*)\leq 2 \sqrt{d} \|S\|_{HS}= 2\sqrt{d}\epsilon$.

If $\| \mathcal{O}\|_1 < 1$, we know that $ \mathcal{E}=\mathbf{1} +
\mathcal{O}$ is
invertible and $ \mathcal{E}^{-1} = \sum_{n\geq 0} (- \mathcal{O})^n$. By
taking the norm, $ \| \mathcal E^{-1} - \mathbf{1}\|_1 \leq \sum_{n\geq 1} \|
\mathcal{O}\|_1^n = f(\epsilon)/(1-f(\epsilon))$.

Channels stabilize $ \mathcal{B}_{sa}( \mathcal{H})$; as $\mathcal E$ is furthermore  invertible, equality of dimension shows that $ \mathcal{E}( \mathcal{B}_{sa}(
\mathcal{H})) = \mathcal{B}_{sa}( \mathcal{H})$ and  $ \mathcal{E}^{-1}(
\mathcal{B}_{sa}( \mathcal{H}))  = \mathcal{B}_{sa}( \mathcal{H})$.

Now, $X$ is positive, so that  $\|X\|_{HS} \leq \sqrt{d} \lambda_M(X)$. This implies
$ \|(\mathcal{E}^{-1}-\mathbf{1})(X) \|_{HS} \leq  \sqrt{d} \lambda_M(X)
f(\epsilon)/(1-f(\epsilon))$, and in turn $
\mathcal{E}^{-1}(X) \geq X - \sqrt{d} \lambda_M(X)
f(\epsilon)/(1-f(\epsilon))\mathbf{1}$. Taking the bottom of the spectrum ends the
proof.

\end{proof}

We are now ready to prove Theorem \ref{reciproque}.

\begin{proof}[ Proof of Theorem \ref{reciproque}.]

We aim at exhibiting a channel $ \mathcal{E}$ and a POVM {\bf Q} such that $
\mathcal{E}(\mathbf{Q}) = \mathbf{P}$ and $Q_e$ has a wider spectrum than $P_e$
for some $e\in E$. Then Lemma \ref{spectrum} proves that {\bf Q} is strictly cleaner than {\bf P}, and in turn that {\bf P} is not clean.

The building blocks are the subspaces supplied by Lemma \ref{determination}.
Since $ \mathcal{H}$ is not determined by $\{\Supp(P_e)\}_{e\in E}$, 
there are two supplementary proper subspaces $\mathcal{V}$ and
$\mathcal{W}$ such that each rank-one element has support included either in $
\mathcal{V}$ or in $ \mathcal{W}$. 

We shall write explicitly several matrices in the forthcoming proof. All of them shall be written on an orthonormal basis $\{e_j\}_{1 \leq j \leq d}$ of $\mathcal H$, chosen so that $\{e_j\}_{1 \leq j \leq \dim(\mathcal V)}$ is a basis of $\mathcal V$. We shall express the matrices as two-by-two block matrices, the blocks corresponding to the subspaces $\mathcal V$ and $\mathcal V^\perp$.

We study separately the following cases:
\begin{enumerate}[(a)]
\item{
\label{id}
 All POVM elements $P_i$ are proportional to the identity, that is $P_i = \mu_i \mathbf{1}$.}
\item{ 
\label{bd}
The POVM is not full-rank, each rank-one element has support either in $\mathcal V$ or in $\mathcal V^\perp$, and all POVM elements are block-diagonal in $\mathcal V$ and $\mathcal V^\perp$. }
\item {
\label{r1p}
Each rank-one element has support either in $\mathcal V$ or $\mathcal V^\perp$, and at least one POVM element is not block-diagonal.}
\item {
\label{r1np}
At least one rank-one element has support neither in $\mathcal V$ nor in $\mathcal V^\perp$. }
\end{enumerate}

As a sanity check, let us prove we did not forget any case. Either our POVM is full-rank, or it is not. In the latter situation, either there is a rank-one element whose support is not included in $\mathcal V$ nor in $\mathcal V^\perp$ -- and we are in case (\ref{r1np}) --, or all rank-one elements are included in $\mathcal V$ or $\mathcal V^\perp$. Then either there is a POVM element that is not block-diagonal -- and we are in case (\ref{r1p}) -- or all POVM elements are block-diagonal -- and we are in case (\ref{bd}). On the other hand, if {\bf P} is full-rank, we may choose the subspaces $\mathcal V$ and $\mathcal W$ any way we like. Notably, if one POVM element $P_i$ is not proportional to the identity, so that it has non-trivial eigenspaces,  we may choose $\mathcal V$ such that $P_i$ is not block-diagonal in $\mathcal V$ and $\mathcal V^\perp$ -- and we are in case (\ref{r1p}). Finally, if on the contrary, all POVM elements are proportional to the identity, we are in case (\ref{id}).

\medskip

\noindent \emph{Case (\ref{id}):} If all POVM elements are of the form $P_i = \mu_i \mathbf{1}$, then,  for any $\mathcal E = \{R_{\alpha}\}$, we have $\mathcal E(P_i) = \sum_{\alpha} R_{\alpha}^* (\mu_i \mathbf 1) R_{\alpha} = \mu_i \sum_{\alpha} R_{\alpha}^* R_{\alpha} = \mu_i \mathbf 1 = P_i$. No channel can change the wholly uninformative measurement {\bf P}.

 On the other hand, many POVMs can be degraded to {\bf P}. Consider for example the POVM given by $Q_1 = \mu_1 |e_1\rangle\langle e_1| + \sum_{j= 2}^d   |e_j\rangle\langle e_j|$ and $Q_i = \mu_i  |e_1\rangle\langle e_1|$ for $i > 1$. Then ${\mathbf Q} \neq \mathbf P$, so that $\mathbf P \not\succ \mathbf Q$. Yet, with $R_{\alpha} = |e_1\rangle\langle e_{\alpha}|$ for $1\leq \alpha \leq d$, we have $\mathcal E(\mathbf Q) = \mathbf P$, and $\mathbf Q \succ \mathbf P$. Hence {\bf P} is not clean.

\medskip

\noindent \emph{Case (\ref{bd}):} Since all rank-one elements are included either in $\mathcal V$ or in $\mathcal V^{\perp}$, we take $\mathcal W = \mathcal V^{\perp}$. We further choose $\mathcal V$ to be the smaller of the two subspaces, that is $\dim(\mathcal V) \leq d/2 \leq \dim(\mathcal W)$. Then there is a matrix 
$A: \mathcal{V}\to \mathcal{W}$ such
that $AA^*=\mathbf{1_{\mathcal V}}$. If all rank-one elements have support in $\mathcal W$, we further impose that at least one of these supports is not included in the kernel of $A$.

We then define $R^*_{ \mathcal{V}} $ and $
R^*_{\mathcal{W}}$  as:
\begin{eqnarray*}
R^*_{ \mathcal{V}}(\epsilon)  = \left[
                       \begin{array}{c|c}     
                       \mathbf{1}_{\mathcal V} & \epsilon A \\
                       \hline
                       0         &  0
                       \end{array} \right],
\\
R^*_{ \mathcal{W}}(\epsilon)  = \left[
                       \begin{array}{c|c}     
                        0 & 0 \\
                       \hline
                       0   &  \mathbf{1}_{\mathcal W}
                       \end{array} \right].
\end{eqnarray*} 
 Their images are respectively
$ \mathcal{V}$ and $ \mathcal{W}$.

From $R_{\mathcal V}(\epsilon)$ and $R_{\mathcal W}(\epsilon)$, we define the channel $\mathcal E_\epsilon= \{R_1(\epsilon), R_2(\epsilon), R_3(\epsilon)\}$:
\begin{eqnarray*}
R_1^*(\epsilon)  = & \sqrt{\frac{\epsilon^2}{1+\epsilon^2}} R^*_{ \mathcal{V}}(
\epsilon)+  \sqrt{\frac{1-\epsilon^2}{1+\epsilon^2}} R^*_{ \mathcal{W}}(
\epsilon) &  = \left[
                       \begin{array}{c|c}     
                       \sqrt{\frac{\epsilon^2}{1+\epsilon^2}} \mathbf{1}_{\mathcal V} &
\sqrt{\frac{\epsilon^4}{1+\epsilon^2}} A \\
                       \hline
                       0         &
\sqrt{\frac{1-\epsilon^2}{1+\epsilon^2}}\mathbf{1}_{\mathcal W}
                       \end{array} \right], \\
R_2^*(\epsilon)  = &  \sqrt{\frac{\epsilon^2}{1+\epsilon^2}} R^*_{ \mathcal{W}}(
\epsilon)  & =  \left[
                       \begin{array}{c|c}     
                       0 & 0 \\
                       \hline
                       0         &
\sqrt{\frac{\epsilon^2}{1+\epsilon^2}}\mathbf{1}_{\mathcal W}
                       \end{array} \right], \\
R_3^*(\epsilon)  =  & \sqrt{\frac{1 - \epsilon^2}{1+\epsilon^2}} R^*_{ \mathcal{V}}(
\epsilon) - \sqrt{\frac{\epsilon^2}{1+\epsilon^2}} R^*_{ \mathcal{W}}(
\epsilon) &  = \left[
                       \begin{array}{c|c}     
                       \sqrt{\frac{1 - \epsilon^2}{1+\epsilon^2}} \mathbf{1}_{\mathcal V} &
\sqrt{\frac{\epsilon^2-\epsilon^4}{1+\epsilon^2}} A \\
                       \hline
                       0         &
-\sqrt{\frac{\epsilon^2}{1+\epsilon^2}}\mathbf{1}_{\mathcal W}
                       \end{array} \right]. \\
\end{eqnarray*}
Since
$AA^*=\mathbf{1}_{\mathcal V}$, we have $\sum_{\alpha} R_{\alpha}^* R_{\alpha} = \mathbf{1}$, hence these matrices $\{R_{\alpha}\}$ define a genuine channel. A few calculations show that the effect of this channel is:
\begin{equation}
\label{effect}
 \mathcal{E}_\epsilon: \left[
                       \begin{array}{c|c}     
                        B & C \\
                       \hline
                        C^*  &  D
                       \end{array} \right]
\to 
                 \left[
                       \begin{array}{c|c}     
                        \frac1{1+\epsilon^2}\left(B +\epsilon (AC^* +
CA^*) + \epsilon^2 ADA^*\right) & 0  \\
                       \hline
                        0  &  D
                       \end{array} \right].
\end{equation} 

Now, for any $w \in \mathcal W$, we have 
\[
\left[\begin{array}{c}-\epsilon A w \\
\hline w \end{array}\right]\left[\begin{array}{c}-\epsilon A w \\
\hline w \end{array}\right]^* = \left[
                       \begin{array}{c|c}     
                        \epsilon^2 Aw w^*A^* & -\epsilon Aw w^*  \\
                       \hline
                        -\epsilon w w^* A^* &  w w^*
                       \end{array} \right],
\]
so that for any sequence of $w_j \in \mathcal W$, the matrix $ \sum_{j,k} \left[
                       \begin{array}{c|c}     
                        \epsilon^2 A w_j w_k^*A^* & -\epsilon A w_j w_k^*  \\
                       \hline
                        -\epsilon w_j w_k^* A^* &  w_j w_k^*
                       \end{array} \right] $ is non-negative. As any non-negative endomorphism $D$ of $\mathcal W$ can be written $\sum_{j,k} w_j w_k^*$ for appropriate $w_j$, we get that for any non-negative $D$, the matrix $\left[
                       \begin{array}{c|c}     
                        \epsilon^2 ADA^* & -\epsilon AD  \\
                       \hline
                        -\epsilon DA^* &  D
                       \end{array} \right]$
is non-negative. Moreover applying equation \eref{effect} yields that its image by $\mathcal E_{\epsilon}$ is 
$\left[
                       \begin{array}{c|c}     
                        0 & 0  \\
                       \hline
                        0  &  D
                       \end{array} \right] $.

Similarly, if $B \in \mathcal B(\mathcal V)$ is non-negative, then  
$\left[
                       \begin{array}{c|c}     
                        (1+ \epsilon^2) B & 0  \\
                       \hline
                        0  & 0
                       \end{array} \right] $ is non-negative and its image by $\mathcal E_{\epsilon}$ is $\left[
                       \begin{array}{c|c}     
                        B & 0  \\
                       \hline
                        0  &  0
                       \end{array} \right] $.

We use these observations to define a map (not a channel) $ \mathcal{F}_{\epsilon}$ on the
block-diagonal matrices:
\begin{equation}
\label{return}
 \mathcal{F}_{\epsilon}:\left[
                       \begin{array}{c|c}     
                        B & 0  \\
                       \hline
                        0  &  D
                       \end{array} \right]
\to
\left[
                       \begin{array}{c|c}     
                        (1+\epsilon^2)B + \epsilon^2 ADA^* & -\epsilon AD \\
                       \hline
                        -\epsilon DA^*  &  D
                       \end{array} \right].
\end{equation}
We get that $ \mathcal{E}_{\epsilon}( \mathcal{F}_{\epsilon}(M))=M$ for all block-diagonal $M$ and
that if furthermore $M\geq 0$ then $ \mathcal{F}_{\epsilon}(M)\geq 0$. 

We now isolate one full-rank element of {\bf P}, say $P_1$. For all $i\neq 1$,
we define $ Q_i(\epsilon) = \mathcal{F}_{\epsilon}(P_i)$. They are non-negative
and fulfil $ \mathcal{E}_{\epsilon}(Q_i(\epsilon)) = P_i$. Define now
$Q_1(\epsilon) = \mathbf{1} - \sum_{i\neq 1} Q_i(\epsilon)$. The closure
relation ensures that $ \mathcal{E}_\epsilon(Q_1(\epsilon))= P_1$.
What's more,  recalling that $\sum_i B_i = \mathbf{1}_{\mathcal V}$ and $\sum_i
D_i = \mathbf{1}_{\mathcal W}$, we obtain:
\begin{eqnarray*}
Q_1( \epsilon) & = \left[
                       \begin{array}{c|c}     
                        \mathbf{1}_{\mathcal V} -(1+\epsilon^2)\sum_{i\neq1}B_i - \epsilon^2
A(\sum_{i\neq1}D_i)A^* & \epsilon A\sum_{i\neq1}D_i \\
                       \hline
                        -\epsilon \sum_{i\neq 1}D_eA^*  &  \mathbf{1}_{\mathcal W} -
\sum_{i\neq1}D_i
                       \end{array} \right]\\
&              = \left[
                       \begin{array}{c|c}     
                        (1+\epsilon^2)B_1 + \epsilon^2 AD_1A^*
-2\epsilon^2\mathbf{1}_{\mathcal V} & \epsilon A(\mathbf{1}_{\mathcal W}- D_1) \\
                       \hline
                        \epsilon(\mathbf{1}_{\mathcal W}-D_1)A^*)  &  D_1
                       \end{array} \right]\\
& \underset{\epsilon\to 0}{\longrightarrow} \left[
                       \begin{array}{c|c}     
                        B_1 & 0  \\
                       \hline
                        0  &  D_1
                       \end{array} \right] \\
& = P_1.
\end{eqnarray*}
Since $P_1$ is positive, this convergence entails the non-negativity of $Q_1(\epsilon)$ for $\epsilon$ small enough.
As
$Q_1(\epsilon)$ has been chosen so that $\sum_e Q_e(\epsilon) =\mathbf{1}$, we
have defined a genuine POVM $\mathbf{Q}(\epsilon)= \{Q_e(\epsilon)\}_{e\in E}$
such that $ \mathcal{E}_{ \epsilon}(\mathbf{Q}(\epsilon)) = \mathbf{P}$, hence $\mathbf Q \succ \mathbf P$.

We end the study of this case by considering a rank-one element $P_i = \mu_i |\psi_i\rangle \langle \psi_i|$ whose support is not in the kernel of $A$. Using formula \eref{return}, if $|\psi_i\rangle \in \mathcal V$, we get $\Tr(Q_i(\epsilon)) = (1+ \epsilon^2) \Tr(P_i) > \Tr(P_i)$, else $|\psi_i\rangle \in \mathcal W$ and we get $\Tr(Q_i(\epsilon)) = \Tr(P_i) + \epsilon^2 \Tr(A |\psi_i\rangle \langle \psi_i| A^*) > \Tr(P_i)$. In both cases, bigger trace implies that the spectrum of $Q_i(\epsilon)$ is wider than that of $P_i$ and Lemma \ref{spectrum} yields $\mathbf P \not\succ \mathbf Q$. So that {\bf P} is not clean.

\medskip

\noindent\emph{Case (\ref{r1p}):}  Since all rank-one elements are included either in $\mathcal V$ or in $\mathcal V^{\perp}$, we take $\mathcal W = \mathcal V^{\perp}$. 

We now define the channel $\mathcal E_{\epsilon}$ through:
\begin{equation*}
R_1(\epsilon) = \epsilon \Pi_{ \mathcal{V}}, \quad R_2(\epsilon) = \epsilon
\Pi_{ \mathcal{W} } = \epsilon \Pi_{ \mathcal{V}^{\bot}}, \quad R_3( \epsilon)
= \sqrt{1-\epsilon^2} \mathbf{1},
\end{equation*} 
where $\Pi$ denotes here orthogonal projection.

For $\epsilon$ small enough, by Lemma \ref{bound}, the channel is invertible as a positive map. We then define $Q_i = \mathcal E_{\epsilon}^{-1}(P_i)$.

Through the formula $\mathcal E_{\epsilon}(Q_i) = P_i$,  we check:
\begin{equation}
\label{formula}
\mathrm{If\,\,\,\,} P_i= \left[
 \begin{array}{c|c} 
 B   & C  \\
 \hline
 C^* & D
 \end{array}
\right], 
\quad \mathrm{then\,\,\,} 
Q_i(\epsilon)= \left[
 \begin{array}{c|c} 
 B   & (1- \epsilon^2)^{-1}C  \\
 \hline
 (1- \epsilon^2)^{-1} C^* & D
 \end{array}
\right]. 
\end{equation} 

The first remark is that the closure relation ensures $\sum Q_{i}(\epsilon) = \mathbf{1}$.

We also notice that, since rank-one elements have support either in $\mathcal V$ or in $\mathcal W = \mathcal V^\perp$, the rank-one elements are block-diagonal and $Q_i(\epsilon) = P_i$ .

We know that at least one POVM element is not block-diagonal. So that there is an $i\in I $ such that $P_i$ is full-rank and $C$ is non-zero (say
$[C]_{j,k}\neq0$). Then, writing $n = \dim(\mathcal V)$,  there
is an $\epsilon_+\in (0,1)$ such that
\begin{eqnarray*}
 [Q_i( \epsilon_+)]_{j,j}[Q_i(
\epsilon_+)]_{n+k,n+k} & = [B]_{j,j}[D]_{k,k} \\
& <
\frac1{1-\epsilon_+^2}|[C]_{j,k}|^2=
[Q_i( \epsilon_+)]_{j,n+k}[Q_i( \epsilon_+)]_{n+k,j} 
\end{eqnarray*}
so that 
we cannot have positivity of $ Q_i( \epsilon_+)$. 

We define the bottom of the spectrum of the images $Q_i$ of the full-rank elements of {\bf P}:
\[
\lambda_m( \epsilon) = \inf_{i | P_i\mathrm{\,\,\, full-rank}}\lambda_m(Q_i(
\epsilon )).
\]

 Equation \eref{formula} implies that the matrix $Q_i(\epsilon)$ is a continuous function of $\epsilon$ for
$\epsilon\in[0,1)$. Hence its spectrum is also a continuous function of $\epsilon$. 
Accordingly, the function $\lambda_m(\epsilon)$  is the minimum of a finite number of continuous function of $\epsilon$, therefore $\lambda_m(\epsilon)$ is continuous. Its value in $0$ is the bottom of the spectrum of the full-rank elements of {\bf P}, that is $\lambda_m( 0) = \inf_{i | P_i\mathrm{\,\,\, full-rank}}\lambda_m(P_i(
\epsilon )) > 0$. Moreover we have just proved that $\lambda_m(\epsilon_+) < 0$. Thus, by the intermediate value Theorem, there is an $\epsilon_+ > \epsilon > 0$ such that $0 < \lambda_m(\epsilon) < \lambda_m(0)$.

 As $\lambda_m(\epsilon) > 0$, the $Q_i(\epsilon) = \mathcal{E}_{\epsilon}(P_i)$ for $P_i$ full-rank are non-negative, and valid POVM elements. Likewise, we already know that $Q_i(\epsilon) = P_i$ is a valid POVM element if $P_i$ is rank-one. Since we have also shown that $\sum Q_i(\epsilon) = \mathbf{1}$, we have proved that $\mathbf Q(\epsilon)$ is a POVM. Furthermore $\mathcal E_{\epsilon} (\mathbf Q(\epsilon)) = \mathbf P$, thus $\mathbf Q(\epsilon) \succ \mathbf P$.

  As $\lambda_m(\epsilon) < \lambda_m(0)$, there is a full-rank element $P_i$ such that $\lambda_m(Q_i(\epsilon)) < \lambda_m(P_i)$. Hence, using Lemma \ref{spectrum}, we get $\mathbf P \not\succ \mathbf Q(\epsilon)$ and {\bf P} is not clean.

Hence $\lambda_m(
\epsilon_+)\leq 0< \lambda_m$. By the intermediate value Theorem, we can find
an $\epsilon_0\in (0,\epsilon_+)$ such that $ \lambda_m(\epsilon_0)=0$. As
$0\leq\lambda_m(\epsilon_0)< \lambda_m$ we have proved that $
\mathbf{Q}(\epsilon_0)\succ\mathbf{P}$ and that {\bf P} is not clean.

\medskip

\noindent \emph{Case (\ref{r1np}):}
As
$
\mathcal{V}$ and $ \mathcal{W}$ are supplementary we may choose a matrix $A\in
M_{\dim(\mathcal V),d-\dim(\mathcal V)}( \mathbb{C})$
such that the non-zero columns of the following block matrix form an
orthogonal (though not orthonormal) basis of $ \mathcal{W}$:
\begin{equation*}
R^*_{ \mathcal{W} } = \left[
\begin{array}{c|c}
0 & A  \\
\hline
0 & \mathbf{1}
\end{array}
\right].
\end{equation*} 
We know that the image of a matrix is spanned
by its columns, so the image of $R^*_{\mathcal W} $ is $ \mathcal{W}$.

We then define
\begin{equation}
\label{B}
B(\epsilon) =\sqrt{\mathbf{1} - \left(\frac{ \epsilon^4}{1- \epsilon^2} +\frac{
\epsilon^2}{(1 - \epsilon^2)^2}\right) AA^*}. 
\end{equation} 
This definition is valid if the matrix under the square root is positive. Now
$\left(\frac{ \epsilon^4}{1- \epsilon^2} +\frac{
\epsilon^2}{(1 - \epsilon^2)^2}\right)$ is going to $0$ with $ \epsilon$, so that 
\begin{equation*}
\lim_{ \epsilon\to 0} \mathbf{1} - \left(\frac{ \epsilon^4}{1- \epsilon^2}
+\frac{
\epsilon^2}{(1 - \epsilon^2)^2}\right) AA^* = \mathbf{1}.
\end{equation*} 
From this we conclude that $\mathbf{1} - \left(\frac{ \epsilon^4}{1-
\epsilon^2} +\frac{
\epsilon^2}{(1 - \epsilon^2)^2}\right) AA^*$ is positive for $\epsilon $ small
enough.

Accordingly, we can define
\begin{equation*}
R^*_{ \mathcal{V} }(\epsilon) = \left[
\begin{array}{c|c}
B(\epsilon) & -\frac{A}{1-\epsilon^2}  \\
\hline
0 & 0
\end{array}
\right].
\end{equation*} 
Notice that the image of $R^*_{\mathcal V}$ is included in $\mathcal V$.

We may now define our channel $\mathcal{E}_{\epsilon}$ by
\begin{eqnarray}
\label{r1}
R_1^*( \epsilon) & =  \epsilon R_{\mathcal V}^*(\epsilon) & = \left[
\begin{array}{c|c}
\epsilon B(\epsilon) & -\frac{\epsilon}{1-\epsilon^2}  A  \\
\hline
0 & 0
\end{array} \right]
\\
\label{r2}
R_2^*( \epsilon) & =
\epsilon R^*_{\mathcal W}
& = \left[
\begin{array}{c|c}
0 & \epsilon A    \\
\hline
0 & \epsilon \mathbf{1}
\end{array}
\right]
\\
\label{r3}
R_3^*( \epsilon) & =  
\sqrt{1 - \epsilon^2} \left(R_{\mathcal V}^*(\epsilon) + R^*_{\mathcal W}\right)
& = \left[
\begin{array}{c|c}
\sqrt{1-\epsilon^2}B(\epsilon) & -\frac{\epsilon^2}{\sqrt{1-\epsilon^2}}  A  \\
\hline
0 & \sqrt{1-\epsilon^2} \mathbf{1}
\end{array} \right].
\end{eqnarray}
Notice that $\sum_{\alpha =1}^3
R_{\alpha}^*(\epsilon)R_{\alpha}(\epsilon) = \mathbf{1}$ so that $ \mathcal{E}(
\epsilon)$ is indeed a channel.

Moreover $
\lim_{ \epsilon\to 0} R_3( \epsilon)  =\mathbf{1}_{ \mathcal{H} }$.
Hence, for $\epsilon$ small enough, $\|R_3 -\mathbf{1}\|_{HS}$ is as small as we want.
So Lemma \ref{channelsidentity} allows us to invert the channel $ \mathcal{E}_\epsilon$ as a  map on $
\mathcal{B}_{sa}( \mathcal{H})$. We
define $\mathbf{Q}(\epsilon)$ by its elements $ Q_i(\epsilon) =
\mathcal{E}^{-1}_{\epsilon}(P_i)$. Let us check that for $ \epsilon$ small enough, $
\mathbf{Q}(\epsilon)$ is still a \emph{bona fide} POVM. 

\smallskip

First the closure relation still holds, as $ \sum_{i\in I} Q_i = \sum_{i\in I}
\mathcal{E}^{-1}(P_i) = \mathcal{E}^{-1}( \mathbf{1})$. Now $
\mathcal{E}(\mathbf{1}) = \sum_{\alpha} R_{\alpha}^*R_{\alpha} = \mathbf{1} $
and taking the inverse $\mathcal{E}^{-1}( \mathbf{1})= \mathbf{1} $. 

Remains then to be shown that all $Q_i(\epsilon)$ are non-negative.

If $P_i$ is full-rank, then its spectrum is included in $[\lambda_m, 1] $, with $\lambda_m > 0$.  If $R_3$ is near
enough of the identity, that is, if $ \epsilon$ is small enough,
the inequality \eref{basspectre} then ensures that $Q_i(\epsilon)$ is still positive.

If $ P_i$ is rank-one  $P_i = \lambda_i|\psi_i\rangle\langle\psi_i|$, then by hypothesis
$|\psi_i\rangle\in \mathcal{V}$ or $|\psi_i\rangle\in \mathcal{W}$. 
As $R_3 $ is invertible for $\epsilon$ small enough, we may consider $|\phi_i\rangle$  non-zero colinear to $(R^*_3( \epsilon ))^{-1} |\psi_i\rangle$.
 Then
$R_3^*(\epsilon)|\phi_i\rangle$ is colinear to $|\psi_i\rangle$, and non-zero.
Notice that $|\phi_i\rangle$ depends on $\epsilon$, even if we drop it in the
notation.
Now 
\begin{eqnarray*}
R_3(\epsilon)^*|\varphi\rangle = \sqrt{1-\epsilon^2}&
\left(R_{\mathcal V}^*(\epsilon)|\varphi\rangle + R_{\mathcal W}^*|\varphi\rangle\right) \\ &
\mathrm{with\,\,\, } R_{\mathcal V}^*(\epsilon)|\phi\rangle \in \mathcal{V} 
\mathrm{\,\,and\,\,\, } R_{\mathcal W}^*|\varphi\rangle \in \mathcal{W}. 
\end{eqnarray*}
Since $\mathcal V$ and $\mathcal W$ are supplementary, the latter equality implies that $R_{\mathcal V}^*(\epsilon) |\varphi\rangle = 0 $ when $R_3^*(\epsilon) |\varphi\rangle \in \mathcal W$ and $R_{\mathcal W}^*(\epsilon) |\varphi\rangle = 0$ when  $R_3^*(\epsilon) |\varphi\rangle \in \mathcal V$.
Definitions (\ref{r1}, \ref{r2}, \ref{r3}) then yield $\mathcal{E}_{\epsilon}(|\phi_i\rangle\langle\phi_i|) = R_{\mathcal W}^* (|\phi_i\rangle\langle\phi_i|) R_{\mathcal W}$ if $|\psi_i\rangle\in \mathcal{W}$ and $\mathcal{E}_{\epsilon}(|\phi_i\rangle\langle\phi_i|) = R^*_{\mathcal V}(\epsilon) (|\phi_i\rangle\langle\phi_i|) R_{\mathcal V}(\epsilon)$
 if
 $|\psi_i\rangle\in \mathcal{V}$. In both cases, the output matrix is of the form $\mathcal{E}_{\epsilon}(|\phi_i\rangle\langle\phi_i|) = C_i |\psi_i\rangle\langle \psi_i|$.
So that  $Q_i(\epsilon) = (\lambda_i/C_i) |\phi_i\rangle\langle\phi_i|$ and is non-negative.

Thus, for $\epsilon$ small enough, all $Q_i(\epsilon)$ are non-negative. We have  proved that $\mathbf{Q}(\epsilon)$ is a POVM. Furthermore, since $\mathcal E_{\epsilon}(\mathbf Q(\epsilon)) = \mathbf P$, we know $\mathbf Q(\epsilon) \succ \mathbf P$.

\smallskip

We must still show that $\mathbf Q(\epsilon)$ is strictly cleaner $\mathbf P$.

By hypothesis, there is a rank-one element $P_i = \lambda_i |\psi_i\rangle\langle\psi_i|$ such that $|\psi_i\rangle \in \mathcal W$ and $|\psi_i\rangle \not\in \mathcal V^\perp$. As above, we write $|\phi_i\rangle$ such that $Q_i(\epsilon) = (\lambda_i/ C_i)|\phi_i\rangle \langle\phi_i|$. We start by proving that $C_i$ is  less than one.
 
 We write $|\phi_i\rangle = v_i + v_i^{\perp }$ with $v_i\in \mathcal{V}$ and $
v_i^{\perp}\in \mathcal{V}^{\perp}$. Since $|\psi_i\rangle \in \mathcal W$, we get:
\[
\mathcal{E}_\epsilon(|\phi_i\rangle\langle\phi_i|)=  R_{\mathcal W}^* (|\phi_i\rangle\langle\phi_i|) R_{\mathcal W} = \left[\begin{array}{c}A
v_i^{\bot} \\
\hline
v_i^{\bot}\end{array}\right]
\left[ \begin{array}{c}A
v_i^{\bot} \\
\hline
v_i^{\bot}\end{array} \right]^*.
\]
As the latter expression is also equal to $C_i |\psi_i\rangle\langle\psi_i|$, we obtain that $C_i$ is the square of the norm of $
\left[ \begin{array}{c}A
v_i^{\bot} \\
\hline
v_i^{\bot}\end{array} \right]$. Therefore $C_i= \|A
v_i^{\bot}\|^2 + \|v_i^{\bot}\|^2$. Notice that  the squared norm of
$|\phi_i\rangle$ is $1 = \|
v_i\|^2 + \|v_i^{\bot}\|^2$.
On the other hand, the image of $|\phi_i\rangle$ by $R^*_{\mathcal V}(\epsilon)$ is $0$, so that $B(\epsilon)v_i-
1/(1-\epsilon^2) A v_i^{\perp} = 0 $. From this we get:
\[
A v_i^{\perp} = (1-\epsilon^2) B(\epsilon) v_i.
\]
Since  $|\psi_i\rangle\not\in
\mathcal{V}^{\bot}$, 
 this equality shows that $v_i\neq 0$.
Now, as $AA^*$ is non-negative we see by (\ref{B}) that $ B(\epsilon)\leq
\mathbf{1}$. A fortiori, for any $\epsilon >0$, we have $(1-\epsilon^2)
B(\epsilon) < \mathbf{1}$. So that:
\[
\|v_i\| > \|(1-\epsilon^2) B(\epsilon) v_i\| = \|A v_i^{\bot}\|.
\]

Thus, we finally obtain 
\[
C_i = \|A
v_i^{\bot}\|^2 + \|v_i^{\bot}\|^2 < \|v_i\|^2 + \|v_i^{\bot}\|^2 = 1.
\]

Hence the biggest eigenvalue of $Q_i(\epsilon) = (\lambda_i / C_i) |\phi_i\rangle\langle \phi_i|$, that is $\lambda_i / C_i$, is strictly bigger than the biggest eigenvalue of $P_i$, that is $\lambda_i$. Lemma \ref{spectrum} then gives $\mathbf P \not\succ \mathbf Q(\epsilon)$, and consequently {\bf P} is not clean.

\end{proof}

\section{Summary for quasi-qubit POVMs and a special case}
\label{summary}

We now gather all our results specific to quasi-qubit POVMs.
\begin{thm}
\label{gather}
A quasi-qubit POVM {\bf P} is clean if and only if it is rank-one or the supports of its
rank-one elements totally determine $ \mathcal{H}$. The algorithm of
section \ref{ialg}
figures out if this is the case. Moreover  if  {\bf Q} is
cleanness-equivalent to {\bf P}, the two POVMs are even unitarily equivalent.
\end{thm}
\begin{proof}
Rank-one POVMs are known to be clean (Theorem \ref{rankone}). If the support
of the rank-one elements of {\bf P} totally determine $ \mathcal{H}$, we
also know that {\bf P} is clean by Theorem \ref{suffcond}. In both cases the
theorems state that for these clean POVMs, cleanness-equivalence is the same as
unitary equivalence.

Conversely, if {\bf P} is neither rank-one nor have rank-one elements that
totally determine $ \mathcal{H}$, then Theorem \ref{reciproque} applies and
{\bf P} is not clean.

Stage (\ref{isr1}) of the algorithm checks whether {\bf P} is rank-one, in which
case it does say that {\bf P} is clean. If {\bf P} is not rank-one, the fact
that it is clean or not depends on the support of its rank-one elements. The
only remaining positive exit of the algorithm is at stage (\ref{cleanexit}) and Lemma
\ref{stage6} proves that in this case the rank-one elements of {\bf P} totally
determine $ \mathcal{H}$. 

Conversely, if the algorithm exits with a negative value, Lemma
\ref{determination} ensures that $ \mathcal{H}$ is not totally determined. 

\end{proof}

  To get further feeling of these conditions we finish by making more
explicit the qubit case,
where the nice thing is that all POVMs are quasi-qubit.

\begin{corol}
\label{qubit}
A POVM {\bf P} for a qubit is clean if and only if it is rank-one or if one can
find three rank-one elements whose supports are two-by-two non-colinear (that
is if they make a projective frame). For these POVMs cleanness-equivalence is
the same as unitary equivalence.
\end{corol}
\begin{proof}
A POVM {\bf P} for a qubit has non-zero elements which can be either of rank
one, or of rank two, as $d=2$. In the latter case, they are full-rank, so we
may apply Theorem \ref{gather} to {\bf P}. 

The only question is when do the supports of the rank-one elements totally
determine $ \mathcal{H}$? They do by Proposition \ref{projtot}  if they include
a projective frame, that is a basis and a vector with all coefficients non-zero
in this basis. As the space is of dimension $2$, this amounts to saying a basis
and a vector non-colinear to any basis vector, that is three vectors two-by-two
non-colinear.

Conversely, if we cannot find a projective frame, then we can find two vectors
$v$ and $w$ such that the support of any rank-one element is $v$ or $w$, and we
can apply Lemma \ref{determination} to obtain that $ \mathcal{H}$ is not
totally determined by the supports of the rank-one elements of {\bf P}. Thus
{\bf P} is not clean.

\end{proof}

\section{Outlook}
\label{outlook}

We have solved the problem of cleanness for quasi-qubit POVMs. The obvious
continuation would be to solve it in the general case. However we do not think
that the condition of Theorem \ref{suffcond} is then necessary. Moreover it must be
made explicit.

The heuristics in Section \ref{heuristics}  suggest that, if the support of $P_i$ are in ``general position''  then it is sufficient for {\bf P} to be clean that $\sum_{e\in E}
d -\dim[\Supp(P_i)] \geq d^2 -1$. Yet, we still need to appropriately define the ``general position'' for general subspaces.

\section*{Acknowledgements}
  We thank Professor d'Ariano for introducing us to the notion of clean POVMs.
 We are also indebted to Sylvain Arlot for many general suggestions on writing. My failure at applying them is the source of all the remaining lack of clarity.

\vfil


\begin{thebibliography}{50}
\bibitem{BAKPW}{F. Buscemi, G. M. d'Ariano, M. Keyl, P. Perinotti, R. F.
Werner, {\em Clean positive operator valued measures} J. Math. Phys. {\bf 46},
082109 (2005)}
\bibitem{LivreKraus}{K. Kraus, {\em States, Effects and Operations} (Springer Verlag, Berlin, 1983)}
\bibitem{projfram}{M. Audin, {\em Geometry} (Springer Verlag, Berlin, 2002)
}
\end{thebibliography}
\end{document}